\documentclass[prl,aps,twocolumn,superscriptaddress,amsmath,amssymb,floatfix,citeautoscript]{revtex4-2}

\usepackage{graphicx}
\usepackage{dcolumn}
\usepackage{bm}
\usepackage{hyperref}

\usepackage{color}
\usepackage{bbm}
\usepackage{amsfonts,amsmath,amssymb,stmaryrd}
\usepackage{bbold}    
\usepackage{todonotes} 
\usepackage[normalem]{ulem} 

\newcommand{\tr}{\text{Tr}}

\newcommand{\rh}{\hat{\rho}}

\newcommand{\half}{\frac{1}{2}}

\newcommand{\Om}{\mathbf{\Omega}}
\newcommand{\Dm}{\hat{\mathbf{\Delta}}}

\graphicspath{{figures/}}

\newcommand{\mfl}[1]{\textcolor{black}{#1}} 
\newcommand{\tom}[1]{\textcolor{black}{#1}} 
\newcommand{\dennis}[1]{\textcolor{black}{#1}} 

\begin{document}


\title{Imaginary-time evolution of  interacting spin systems in the truncated Wigner approximation
}

\author{Tom Schlegel}
\author{Dennis Breu}
\author{Michael Fleischhauer}
\affiliation{Department of Physics and Research Center OPTIMAS, RPTU University Kaiserslautern-Landau, D-67663 Kaiserslautern, Germany}

\date{\today}

\begin{abstract}
We present a semiclassical phase-space method to calculate thermal and ground states of large interacting spin systems. To this end, we 
extend the recently developed truncated Wigner approximation for spins (TWA) to the imaginary time, termed iTWA. The evolution of the canonical density matrix in imaginary time is mapped to a partial differential equation of its Wigner function. Truncation at the Fokker-Planck level leads to 
a set of stochastic differential equations, which can be efficiently simulated \mfl{even for large systems}. 
\mfl{We show that for general Ising Hamiltonians the  approximation becomes exact for large imaginary times subject only to sampling errors. Thus the iTWA is ideal to determine the ground state of spin glasses or to find solutions to quadratic unconstrained \dennis{binary} optimization problems \dennis{(QUBO)} on a controlled approximation level.
We illustrate this for MaxCut on random, unweighted 3-regular graphs, encoded in an anti-ferromagnetic Ising Hamiltonian, for which finding the exact ground state and even approximations to it beyond a certain accuracy is know to be NP hard.} 
 Furthermore,
in order to assess the \mfl{quality} of the method \mfl{also for general spin models}, we analyze the ground-state quantum phase transition of the transverse-field Ising model in one and two spatial dimensions, finding \mfl{reasonably} good agreement with the exact behavior.
\end{abstract}

\maketitle


\paragraph{Introduction -- }

Calculating the thermal or ground state of large interacting spin systems is an important problem in many-body physics. In the absence of frustration this task can be efficiently tackled using quantum Monte Carlo (QMC) techniques \cite{suzuki1976relationship,suzuki1977monte,foulkes2001quantum}, mapping the quantum system to an equivalent classical one. For frustrated models, on the other hand, this can lead to configurations with a negative Boltzmann weight and an associated exponential growth of statistical errors that render QMC simulations unreliable \cite{Troyer-PRL-2005}. A generic method to efficiently determine the ground state of frustrated spin models is believed to not exist as it would 
allow to solve an NP complete problem efficiently and thus
imply NP=P. 

Here we propose and analyze an approximate semiclassical approach to calculate the finite-temperature and ground state of a 
spin Hamiltonian. It is 
based on a mapping of the canonical density operator to phase space. To this end we extend the truncated Wigner approximation (TWA) for spins in continuous phase space \cite{MinkPRR2022,mink2023collective} to imaginary time
(iTWA). 
We show that the evolution of the canonical density matrix in imaginary time can be mapped to a partial differential equation for the Wigner function, which represents the density matrix in phase space. Truncating higher-order derivatives and 
exploiting the non-uniqueness of the mapping between Hilbert and phase space
as well as a gauge freedom 
then leads to a Fokker-Planck equation \mfl{(FPE)}, which is equivalent to a set of stochastic differential equations (SDE) that can be simulated efficiently. 
Like its real-time counterpart, the iTWA is a semiclassical approach that goes beyond a mean-field description and takes into account quantum effects. 
For purely unitary dynamics, they are included in the real-time 
TWA 
only by sampling over some initial distribution \cite{DTWA,MinkPRR2022}. This 
restricts their applicability in the absence of dissipati\mfl{on}
to short times~\cite{pucci2016simulation,wurtz2018cluster,zhu-NJP-2019,czischek2018quenches}. In the imaginary-time method on the other hand,  quantum fluctuations are incorporated also through additional noise terms.
\mfl{We here show that for general Ising-type Hamiltonians, the iTWA becomes exact for large imaginary times, subject only to sampling errors and thus can be applied e.g. to find the ground state of Ising spin glasses or to solve NP hard optimization problems on a controlled approximation level.}

To illustrate the power of the method we apply it to a spin system with
frustrated anti-ferromagnetic Ising interaction on a random unweighted 3-regular graph, which encodes 
MaxCut. Determining the ground state 
and even its approximation \mfl{with an error less than $1/17 \approx 5.9 \%$ in energy} is known to be NP hard 
\cite{bermankarpinski1997,10.1145/502090.502098},
\mfl{while the best classical algorithms find solutions above $6.8\%$ with polynomial effort}
\cite{feige2002improved,halperin2004max,khot2007optimal}.
Benchmarking the iTWA with exact diagonalisation results for randomly chosen 3-regular unweighted graphs with $N=22$ spins, we show that the method accurately predicts the ground state.
Considering large graphs with \tom{$N \le 216$} we
show that the method gives very good approximations to the 
ground state, determined with the \mfl{commercial}  program GUROBI \cite{gurobi}, \mfl{limited only by finite sampling errors. The NP-hardness of the problem manifests itself in an exponential scaling of the required number of trajectories.}

In order to assess the ability of the iTWA to describe quantum effects \mfl{also} in spin systems \mfl{with other couplings}, we then investigate the quantum phase transition (QPT) in the  transverse-field Ising model in 1D and 2D. We show that the iTWA simulations correctly reproduce the quantum universal behavior, following the quantum-classical correspondence
\cite{sachdev1999quantum}.

\paragraph{TWA for spins -- }

The truncated Wigner approximation was originally developed to simulate the many-body dynamics of spins in a discrete phase space \cite{DTWA,zhu-NJP-2019}, termed 
DTWA, and has been applied to several interaction problems \cite{perlin-PRB-2020,Czischek-QSciT-2018,Khasseh-PRB-2020,sundar2019dtwaBenchmark,kunimi2021dtwaBenchmark,huber-SciPost-2021,huber2021realistic}. In DTWA quantum effects are 
taken into account only by  sampling of the initial state. To increase the precision of the method, extensions and refinements have been proposed \cite{wurtz2018cluster,Czischek-QSciT-2018,huber-SciPost-2021,Singh2021}. 
The discrete method has then been 
embedded into a continuous phase-space approach in \cite{MinkPRR2022,mink2023collective}, allowing a systematic derivation of approximate equations of motion and the incorporation of reservoir interactions. The continuous method was then successfully applied to collective dynamical phenomena in interacting spin systems such as superradiance \cite{mink2023collective,tebbenjohannsPRA2024,spahn2026motioninduceddirectionalitycollectiveemission}. 

The mapping between Hilbert space and phase space is based upon an expansion of all operators $\hat A$ of a system of $N$ spin-1/2 systems onto an (overcomplete) basis  of phase-point operators 
$\hat{\mathbf{\Delta}}(\mathbf{\Omega})=\prod_{j=1}^N \hat{\Delta}_j(\Omega)$, where every spin is described by two angles $\theta$ and~$\phi$,
\begin{align*}
    \hat{\Delta}(\theta, \phi)
    =& \half
    \begin{pmatrix}
        1 + \sqrt{3} \cos \theta && \sqrt{3} e^{-i \phi} \sin \theta\\
        \sqrt{3} e^{i \phi} \sin \theta && 1 - \sqrt{3} \cos \theta
    \end{pmatrix}, \label{eq:originKernel}
\end{align*}
with $\Omega=(\theta,\phi)$ and $\mathbf{\Omega}=(\Omega_1,\Omega_2,\dots, \Omega_N)$. $\hat A= \int\! d\Om\, {\cal A}(\Om)\,  \hat{\mathbf{\Delta}}(\mathbf{\Omega})$ where the integration $d\Om=\prod_{j=1}^N d\Omega_j$ is over the angles $\theta_j,\phi_j$, i.e. $\int d\Omega = \frac{1}{2\pi}\int_0^{2\pi}\!\!\! d\phi \int_0^\pi \!\! d\theta\sin\theta.$ 
${\cal A}(\Om)$, which is the phase-space representation of the operator $\hat A$ is called Weyl symbol and obeys ${\cal A}(\Om) = \tr\left\{\hat{A}\, \Dm(\Om)\right\}$. Of particular interest is the Weyl symbol of the density operator $\rh$, called Wigner function $W(\Om)$. 

In  TWA  an equation of motion for $W(\Om,t)$ is obtained from the 
master equation $\partial_t \rh = {\cal L} \rh$, where ${\cal L}\rh = -i [\hat H,\rh] +{\cal L}_\textrm{diss} \rh$, where $\hat H$ is the Hamiltonian and $
{\cal L}_\textrm{diss}$ the dissipator, by using correspondence rules 
\begin{align*}
    \partial_t\rh = \int\!\!d\Om\,  \partial_t W(\Om,t) \Dm(\Om)= \int \!\!d\Om\,  W(\Om,t) \bigl[{\cal L} \Dm(\Om)\bigr].
\end{align*}
The action of ${\cal L}$ on the phase point operators corresponds to evaluating products of Pauli spin operators and phase point operators.
Such products can however be expressed in terms of a complete set formed by four operators
$    \hat \Delta,\enspace \partial_\theta  \hat \Delta, \enspace  \partial_\phi  \hat \Delta, \enspace\textrm{and}\enspace  \partial^2_\phi  \hat \Delta, $ or $  \partial^2_\theta \hat \Delta,$
i.e. by the phase point operator itself and its partial derivatives with respect to $\theta$ and $\phi$. Partial integration in the Master equation than leads to a partial differential equation for the Wigner function, which upon approximate truncation to the \mfl{FPE} level can be efficiently simulated by equivalent stochastic differential equations (SDEs) \cite{gardiner2004quantum}.
The integration in  phase space $\int d\Omega$ over the surface of a sphere is conveniently transformed into integration over a torus $\int_0^\pi\!\! d\theta \, \int_0^{2\pi}\!\! d\phi$ by introducing a flattened Wigner function $\chi(\theta,\phi)= W(\theta,\phi) \sin\theta/2\pi$.
The non-uniqueness of the correspondence rules and the gauge freedom of the Wigner function
(see Ref.\cite{MinkPRR2022,Hartmann}) can be used to optimize the SDEs.

\paragraph{Imaginary-time evolution -- }
%
Our aim is to find an approximate expression for the Wigner function of the canonical density matrix at inverse temperature~$\tau$
\begin{equation}
\label{eq:gibbs}
    \hat \rho(\tau) = \frac{1}{Z(\tau)} e^{-\tau \hat H}
\end{equation}
where $\hat H$ is the Hamiltonian and $Z(\tau) = \textrm{Tr}\{e^{-\tau \hat H}\}$.
Since typical spin Hamiltonians are bounded from below and above, the density matrix becomes the fully mixed state in the infinite temperature limit $\tau = 1/k_B T \to 0$. Thus, the state \mfl{obeys} the evolution equation 
\begin{equation}
    \label{eq:Density-EOM}
    \partial_\tau \hat \rho = -\bigl(\hat H - \langle \hat H\rangle\bigr) \hat \rho = -\frac{1}{2}\bigl\{\hat H,\hat \rho\bigr\} + \langle \hat H\rangle \hat \rho.
\end{equation}
Applying the correspondence rules between Hilbert and phase space one finds in contrast to the real-time case, that the equation of motion of the (flattened) Wigner function is not of generalized \mfl{FPE} type, 
but now also contains a term proportional to $\chi(\Om,\tau)$. Separating the contribution form 
$\langle \hat H(\tau)\rangle$, i.e.
$\chi = \Tilde{\chi} \exp\bigl\{\int_0^\tau \!d{\tau^\prime} \langle \hat H\rangle(\tau^\prime) \bigr\} = \tilde \chi / \zeta(\tau)$  yields a partial differential equation for~$\tilde \chi$
\begin{eqnarray}
    \partial_\tau \tilde\chi(\Om,\tau) &=& -{\cal H}(\Om)\tilde \chi(\Om,\tau) -
    \sum_j\frac{\partial}{\partial x_j} \Bigl[A_j(\Om) \tilde\chi(\Om,\tau)\Bigr] \nonumber \\ 
    && +\,\frac{1}{2}\sum_{jl} \frac{\partial^2}{\partial x_j\partial x_l} \Bigl[D_{jl}(\Om) \tilde \chi(\Om,\tau) \Bigr]+\cdots\label{eq:PDE}
\end{eqnarray}
where ${\cal H}(\Om)$ is the Weyl symbol of the Hamiltonian, and $x_k\in\{\theta_j,\phi_j\}$. $\zeta(\tau)=Z(\tau)/Z(0)$ is the ratio of the partition functions at imaginary time $\tau$ and $0$.
Exploiting the degrees of freedom resulting from the non-uniqueness of the correspondence rules as well as the gauge freedom of the Wigner function,
truncation at second order derivatives, and approximating the coefficient matrix by a positive definite one, i.e. $\mathbf{D} \to  \mathbf{B}\cdot \mathbf{B}^\top$ leads to a proper 
\mfl{FPE}
with an additional term proportional to $-{\cal H}(\Om) \tilde\chi$. As will be shown in detail in the Supplementary the solution is equivalent to solving a set of SDEs:
\begin{equation*}
    dx_j(\tau) = A_{j} (\Om)\, d\tau + \sum_l B_{jl}(\Om)\,  dW_l(\tau),
\end{equation*}
where $x_j=(\theta_j,\phi_j)$ and $dW_l$ is the increment of a standard Wiener process. The initial values $x_j(0)$ are drawn from the distribution $\tilde\chi(0)=\chi(0)$. 
Expectation values of operators in the canonical state at inverse temperature~$\tau$ can then be obtained from
\begin{eqnarray}
   && \langle \hat A\rangle(\tau) = \frac{\displaystyle{\overline{{\cal A}(\Om(\tau)) e^{-\int_0^\tau \!\! d\tau^\prime {\cal  H}(\Om(\tau^\prime))} }}}{\displaystyle{\overline{e^{-\int_0^{\tau} \!\! d\tau^{\prime} {\cal  H}(\Om(\tau^{\prime}))} }}}.\label{eq:average-generalized}
\end{eqnarray}
Here the overbar denotes stochastic averaging over the trajectories following the SDE.

Since ${\cal H(\tau)}$ may be negative, the averaging over trajectories can lead to large fluctuations. This can be alleviated by replacing
${\cal H}(\tau) \to {\cal H}(\tau) - E(\tau)$ in numerator and denominator of eq.\eqref{eq:average-generalized}, with appropriately chosen value~$E(\tau)$, \mfl{e.g. $E(\tau) = \overline{{\cal H}(\tau)}$}.

\paragraph{AF Ising models 
--}
%
Finding the ground state of (anti-ferromagnetic, AF) Ising Hamiltonians on general graphs is an important problem in statistical mechanics as it 
is closely related \mfl{to spin glass physics \cite{edwards1975theory,parisi1979infinite,binder1986spin} and} to the solution of 
combinatorical optimization problems \cite{kirkpatrick1983optimization,du1998handbook} 
such as Maximum Independent Set or MaxCut,
as well as machine learning \cite{bishop2006pattern}. 
This task is in general NP hard, \mfl{and includes some NP-complete problems \cite{lucas2014ising}}. 
Thus, a general numerical method to calculate ground states in an efficient way is believed to not exist as this would imply NP=P. 
\mfl{We now show that the iTWA is able to approximate the ground state of \textit{any} Ising Hamiltonian to an arbitrary accuracy level, limited only by finite sampling. Although the iTWA does not change the algorithmic complexity of the problem, reflected in an exponential scaling of the number of trajectories needed to reach accuracy above the NP-hardness threshold, the approximation level is well controlled.}

\mfl{A general Ising Hamiltonian on a graph ${\cal G}$ reads
\begin{equation}
    \label{eq:random-Ising}
    \hat H =  \frac{1}{2}\sum_{i,j\in {\cal G} } J_{ij} \hat\sigma_i^z\, \hat \sigma_j^z,
\end{equation}
where the sum extends over all spins connected by the edges of ${\cal G}$. Since $\hat H$ commutes with all $\hat \sigma^z_j$, the ground and finite temperature states factorize in eigenstates of $\hat\sigma^z$. Thus it is sufficient to consider the Wigner function integrated over the azimutal angles, $\tilde\chi_\theta =  \prod_j \int\! d\phi_j\, \tilde\chi(\mathbf{\Om})$. \mfl{Then,} using $\partial_\phi^2 \hat \Delta$ in the correspondence rules\tom{,} 
the following drift vector and diffusion matrix \tom{are the only remaining terms} 
%
\begin{eqnarray*}
    A_{\theta_i} &=& -\sum_{i,j\in{\cal G}} J_{ij}\Bigl(\frac{2}{\sin\theta_i}-3\sin\theta_i\Bigr) \cos\theta_j,  \\
     D_{\theta_i\theta_j} &=& -\frac{J_{ij}}{3}\Bigl(\frac{2}{\sin\theta_i}-3\sin\theta_i\Bigr)\Bigl(\frac{2}{\sin\theta_j}-3\sin\theta_j\Bigr).\,\,
\end{eqnarray*}
The diffusion matrix is in general not positive and thus needs to be neglected for a mapping to SDEs. However, the deterministic imaginary-time evolution leads to the asymptotics 
$\sin^2 \theta_j  \rightarrow  \frac{2}{3}$ for $\tau\to\infty$,
corresponding to eigenstates of $\hat \sigma_j^z$ with eigenvalues $\pm 1$. Thus for $\tau \to \infty$ the diffusion matrix vanishes $\vert D_{\theta_i\theta_j}\vert \to 0$ and the iTWA becomes exact!
}
\tom{In addition, even for small $\tau$ the contribution of the diffusion matrix can be kept small $\vert D_{\theta_i\theta_j}\vert < 1$ by using a suitable sampling of the fully-mixed state} \mfl{(see Supplementary Material).}

In the following, we investigate 
an AF Ising Hamiltonian with uniform coupling strength $J$ between nodes of a random 3-regular graph ${\cal G}$, 
in which every node (spin) is connected by exactly three edges to other nodes (spins) \mfl{(As a different example random Erd\"os-Renyi graphs are discussed in the Supplementary material.)}. For such graphs, finding the ground state 
or even approximations to it below a relative error $\Delta \epsilon=1/17$
is NP-hard because Max-Cut is known to be NP-hard 
\cite{10.1145/502090.502098} and there exists a one to one mapping between both problems \cite{PhysRevResearch.6.023294}.
The eigenstate with highest energy is obviously the state with all spins pointing the same direction with $E_\textrm{max}/J = 3 N/2$. Note that $N$ is even for a 3-regular graph. Since in the unweighted graph every node has three equivalent edges, all other possible energies are $E_n/J= 3 N/2 - (2 n +1)$, with $n$ being some non-negative integer
upper bounded by $3N/2-1$. 

In the following we calculate the finite temperature state of the model with the iTWA.
As noted above, the diffusion matrix is in general not positive and thus must be neglected\mfl{, an approximation which becomes however exact in the limit $\tau\to\infty$.} The stochastic equations for eq.~\eqref{eq:random-Ising} are thus ordinary \mfl{nonlinear} differential equations. 
\begin{align}
    \frac{d \theta_i}{d \tau} = -J \Bigl(\frac{2}{\sin\theta_i}-3\sin\theta_i\Bigr) \sum_{i,j\in{\cal G}} \cos\theta_j 
    \nonumber
\end{align}
and can be efficiently solved using GPUs.

\begin{figure}[h]
    \centering
        \includegraphics[width=0.48\textwidth]{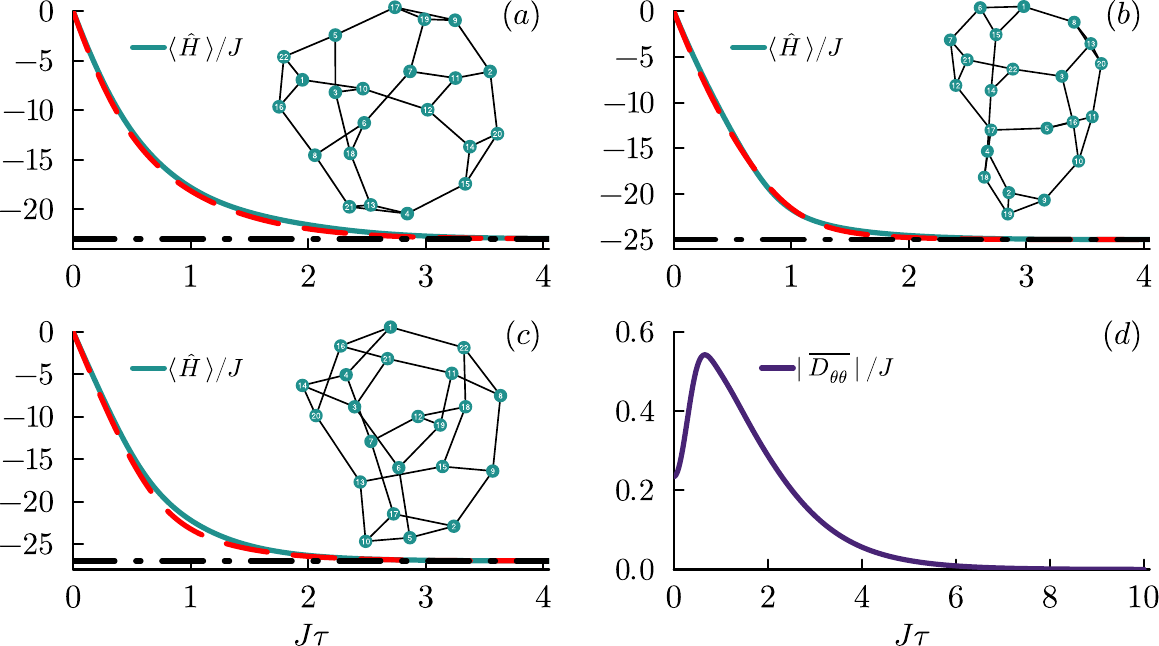}
        \caption{\mfl{(a) to (c): } Average energy $\langle \hat H\rangle$ of the Gibbs state of the AF Ising Hamiltonian, eq.~\eqref{eq:random-Ising}, on randomly chosen 3-regular graphs with $N=22$ nodes over \tom{$N_\mathrm{traj} = 10^5$} trajectories as function of inverse temperature $\tau = 1/k_B T$. Shown are iTWA (solid, green) and exact results (ED, dashed, red) as well as the exact ground state energies $E_0=-23J, -25J, -27J$ (dot-dashed, black) together with  illustrations of the 3-regular-graphs. \tom{(d) Norm $\lVert \cdot\rVert_2 = \sqrt{\sum_j \lvert a_j\rvert^2}$ of the (non-positive) diffusion matrix $D_{\theta_i\theta_j}$ for the graph in (c).  
        The diffusion matrix vanishes as $\tau \to \infty$ and remains small for 
        suitable initial samplings.}}
        \label{fig:Ising-N22-3ru}
\end{figure}

In Fig.\ref{fig:Ising-N22-3ru} we have plotted the average energy~$\langle \hat H\rangle = \textrm{Tr}\{\hat \rho\,  \hat H\}$ for the Gibbs state, eq.~\eqref{eq:gibbs}, for a few randomly chosen 3-regular graphs with $N=22$ nodes. Shown are the iTWA results and values obtained by exact diagonalization (ED). One recognizes very good agreement between iTWA and exact results over the whole range of inverse temperatures and  perfect agreement of the ground-state energies, which due to the gapfulness of the Hamiltonian is reached for $J \tau \geq 3 $.
\mfl{Also shown is the norm of the non-positive diffusion matrix $D_{\theta_i\theta_j}$ which approaches zero in the same limit.}

\begin{figure}[h]
    \centering
    \includegraphics[width=0.48\textwidth]{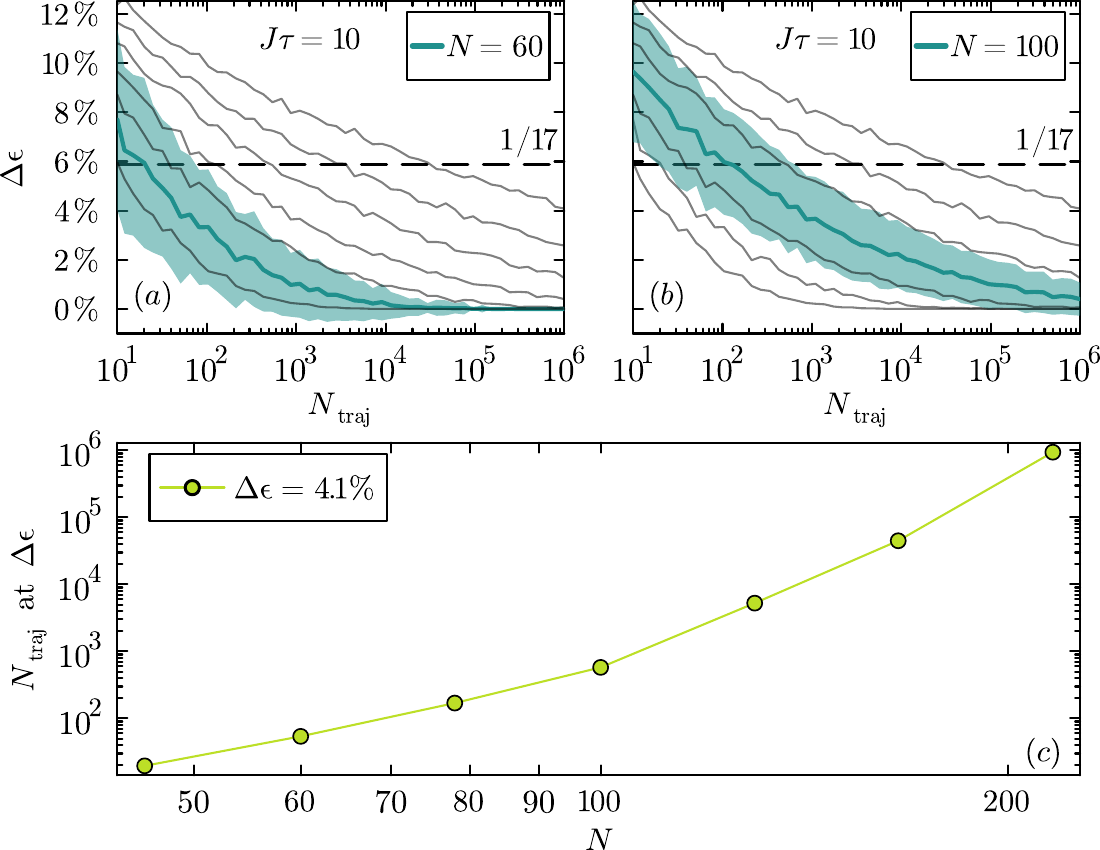}
    \caption{
    \tom{(a)+(b):} Mean relative error \tom{$\Delta \epsilon = (E_0 - \langle \hat H \rangle) / E_0$}  simulated via iTWA at an annealing time of $J\tau=10$ to the exact ground state energy $E_0$ estimated by Gurobi as a function of the number of trajectories $N_\mathrm{traj}$ for \tom{ different $N=46, 60, 78, 100, 130, 166, 216$ (thin lines). The data are averaged over 200 randomly chosen 3-regular graphs.  Highlighted are the system sizes $N=60$ and $N=100$ with the standard deviation as error bar}. Note that the underlying distribution is not normal and is \mfl{lower} bounded \mfl{by $E_0$} for all trajectories.
    \mfl{(c): Minimum number of trajectories $N_\textrm{traj}$ required to approach the ground state with relative error $\Delta \epsilon=4.1\%$ as function of system size~$N$.}}
    \label{fig:Ising-large-3ru}
\end{figure}

We now illustrate the power of the iTWA to find (approximate) solutions of the ground state of eq.~\eqref{eq:random-Ising} on large graph\tom{s} with \tom{$N=\mathcal{O}(100)$}, which cannot be solved anymore by ED. 
\dennis{To estimate the ground-state we use the state of the art classical solver GUROBI \cite{gurobi}.} 
Being a stochastic algorithm (note that even in the absence of a diffusion term the initial values must be sampled) the quality of its predictions depends on the number of trajectories. In 
Fig.~\ref{fig:Ising-large-3ru}(a+b) 
we plot the relative deviation of the average energy $\langle \hat H\rangle$ for the Gibbs state at $J\tau =10$ (which is very close to the ground state) as a function of the number of trajectories $N_\textrm{traj}$ \mfl{for different $N$.} 

\mfl{As the iTWA is exact for Ising Hamiltonians for $\tau\to\infty$ the quality of the approximation of the ground state is only limited by finite-sampling errors. In Fig.~\ref{fig:Ising-large-3ru}(c) we have shown the system-size scaling of the number of trajectories $N_\textrm{traj}$ needed to approach the ground state within an error $\Delta \epsilon =4.1\%$ of the trajectory averaged energy $\langle \hat H\rangle$ at $J\tau =10$ for the graphs shown in Fig.~\ref{fig:Ising-large-3ru}(a+b). The scaling becomes exponential, as expected from the NP-hardness of finding the ground state.}

\paragraph{Transverse-field Ising model -- }

\mfl{Another} application of the iTWA is the calculation of the ground state of \mfl{more general} spin models. While for frustrated spin systems with \mfl{competing interactions and} high ground-state entanglement we do not expect that iTWA simulations yield accurate results, it should work for non-frustrated models. To check this, we consider the transverse-field \mfl{ferromagnetic} Ising model \tom{(TFIM)} with nearest neighbor interactions on a 1D chain or a 2D square lattice.    
\begin{equation}
    \label{eq:Transversse-field-Ising}
    \hat H = - h\sum_j\hat \sigma_j^x - \frac{J}{2} \sum_{\langle ij\rangle } \hat\sigma_i^z\, \hat \sigma_j^z
\end{equation}
The Hamiltonian has a $Z_2$ symmetry $\hat \sigma_z \to - \hat \sigma_z$ and the system undergoes a quantum phase transition (QPT) \cite{sondhi1997continuous,sachdev1999quantum} in the ground state \mfl{at a critical } ratio 
\mfl{of} \tom{$h/J$}, 
where this symmetry is spontaneously broken. As first shown by Suzuki \cite{suzuki1976relationship}, the critical behavior of the quantum  model in $d$ spatial dimensions is identical to that of the thermal transition of the classical model in $d+1$ dimensions. Furthermore the 1D model can be solved exactly by Wigner-Jordan and Bogoliubov transformations.

Since we expect quantum noise to be relevant in the QPT, we 
exploit the freedom in the choice of correspondence rules to minimize the approximations needed to bring the equation of motion of the Wigner function to Fokker-Planck form, without having to neglect the diffusion contributions \mfl{altogether. Details are given in the Supplementary material.}

\begin{figure}[h]
    \centering
        \includegraphics[width=0.4\textwidth]{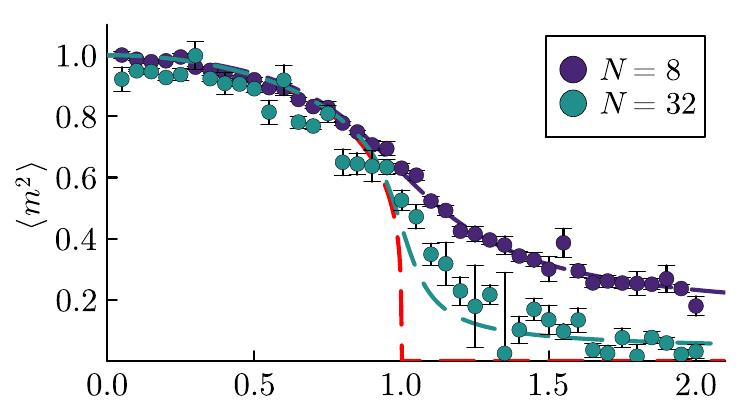}
        \includegraphics[width=0.4\textwidth]{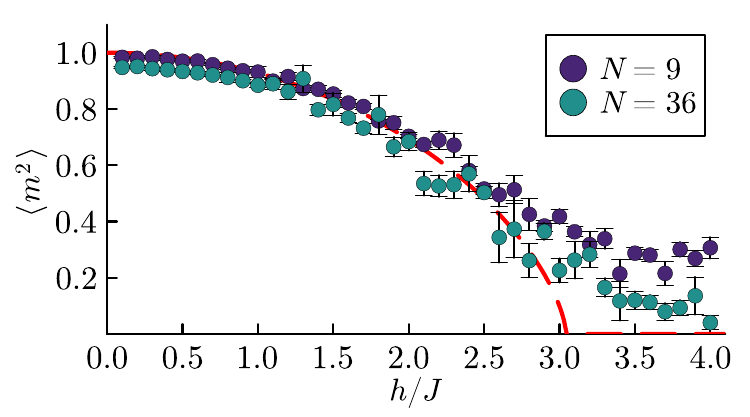}
        \caption{Total squared magnetization $\langle m^2\rangle$ for the 1D (top) and 2D (bottom) TFIM for different number of spins with exact finite-size (green and blue dashed lines) in 1D and infinite-size solutions (red dashed line) in 1D and 2D. The stationary values were determined by averaging $\langle m^2\rangle$ over $\tau$. In 1D in the interval $\tau=3/J-5/J$ and in 2D in the interval $\tau = 1/J-3/J$. For the error bars the leave-one-out jackknife standard deviation \cite{dca15e5b-b3f7-3417-8555-955fe36eb045} was used to account for bias in single trajectories (1D: $N_\mathrm{traj} = 480\cdot10^3$, 2D: $N_\mathrm{traj} = 600\cdot10^3$).}
        \label{fig:QPT}
\end{figure}

We have simulated the thermal state $\hat\rho(\tau)$ for sufficiently large values of $\tau$, which is a very good approximation to the ground state, for the TFIM for different values of $h/J$. The ground state undergoes a quantum phase transition at $h/J=1$ in 1D and $h/J \approx 3.044$ in 2D, which corresponds to the universality class of the corresponding classical models in $d+1$ dimensions \cite{suzuki1976relationship,sachdev1999quantum}. In Fig.\ref{fig:QPT} we have plotted the order parameter $\langle m^2\rangle = (N + \sum_{j\neq i} \langle \sigma_i^z \sigma_j^z \rangle)/N^2$ obtained from iTWA for a 1D chain and a quadratic 2D lattice, and compared it to exact results in the thermodynamic limit (dashed red lines). The 1D TFIM can be solved exactly also for finite systems
(dashed green and blue lines).
One recognizes that the iTWA reproduces the exact results, \mfl{albeit with sizeable fluctuations}, and thus is \mfl{in principle} capable to describe a quantum phase transition.

%

\paragraph{Summary -- }

Determining the ground or thermal state of large ensembles of interacting spin-$1/2$ systems is an important, and in general numerically challenging task, in particular in the presence of frustration. We presented a semiclassical method that allows to approximately calculate these states employing stochastic simulations in imaginary time. 
It goes beyond mean-field approximations by taking into account quantum fluctuations through (i) sampling of initial states and (ii) stochastic terms in the imaginary-time equation of motion. \mfl{The iTWA is an exact method to determine the ground state of general Ising Hamiltonians with controlled approximation ratio, given by finite-sampling errors. Thus it can be used to study the ground state of Ising spin glasses and to approximately solve QUBO problems.}

 To illustrate potential applications we considered the AF Ising model on random 3-regular graphs. Finding the ground state of such models is known to be NP hard.
 For small systems, tractable by exact diagonalization, we found very good agreement between iTWA and exact results. For \tom{large graphs with $N\leq216$} nodes we demonstrated that iTWA simulations approach the estimate of the ground state obtained from the \dennis{state of the art classical solver GUROBI \cite{gurobi}}
 with very high accuracy, \mfl{limited only by finite sampling. We showed that the system-size scaling of the required number of trajectories becomes exponential 
 reflecting the NP-hardness of finding the ground state.
Thus, while the iTWA does not provide efficient solutions to an NP-hard problem, it is an computationally inexpensive approximation method. } 
 Considering the transverse-field Ising model to \mfl{test} its \mfl{performance for general spin models}, we showed that the method \mfl{does capture} certain quantum phase transitions \mfl{albeit with larger numerical fluctuations}.
 While we do not expect the method to give good results for problems with
highly entangled ground states, such as spin liquids, our results show that the iTWA is a \mfl{numerically inexpensive} tool to simulate the imaginary-time dynamics of  non-trivial many-body spin problems.

\

 \paragraph*{Acknowledgements:}  
The authors thank Jens Hartmann for fruitful discussions. Financial support from the DFG through SFB TR 185, Project No. 277625399, is gratefully acknowledged. The authors also thank the Allianz für Hochleistungsrechnen (AHRP) for giving us access to the “Elwetritsch” HPC Cluster through the AHRP project TWEAQING \dennis{as well as GUROBI for providing an increased academic license.}

 \paragraph*{Authors contributions:}
T.S. developed the analytic approach together with M.F. He performed the numerical simulations of the QPT in the transverse-field Ising model, and, with support of D.B., for the 3-regular graphs. D.B. performed the ED simulations for the 3-regular graphs. M.F. conceived and supervised the project. All authors discussed the results and contributed to the final version of the manuscript.

\vfill
\bibliographystyle{apsrev4-2}
\bibliography{iTWA.bib}

\section*{Supplementary}


\subsection{Solution of eq.~\eqref{eq:PDE}}

Let us consider a quasi-probability function $W(x,\tau)$  of a single variable $x$ that obeys a
Fokker-Planck type equation with additional linear term similar to eq.~\eqref{eq:PDE}:
\begin{equation}
    \begin{split}
    \partial_\tau W(x,\tau) = &- V(x) W(x,\tau) -\frac{\partial}{\partial x} A(x)\, W(x,\tau)\\
   & +\,\frac{1}{2}\frac{\partial^2}{\partial x^2} D(x)\, W(x,\tau),
    \end{split}
   \label{eq:FPE-W}
\end{equation}
with initial value $W(x,0)=W_0(x)$. 
Here $A(x)$ is a drift coefficient and $D(x)=B^2(x)$ is a positive diffusion coefficient. 
We now show that expectation values taken with the quasi-probability $W(x,\tau)$ can be expressed as
\begin{align*}
    \int_{-\infty}^\infty \!\!\! dx\,\, f(x) \, W(x,\tau) &= \overline{f(x(\tau))\, e^{-\int_0^\tau \!\! d\tau^\prime V(x(\tau^\prime))}}\\ &\equiv \overline{g\bigl(x(\tau),\tau\bigr)}
\end{align*}
where $x(\tau)$ is a stochastic variable that is the solution of the Ito SDE
\begin{equation*}
    dx(\tau) = A\bigl(x\bigr) \, d\tau + B\bigl(x\bigr)\, dw(\tau)
\end{equation*}
with initial condition $x(0)=x_0$,
where $dw(\tau)$ is a standard Wiener increment~\cite{StochasticMethods}. 
Using Ito's formula one finds
\begin{align*}
    dg(x,\tau) = &\left[-V(x) g + A(x) \frac{\partial g}{\partial x} + \frac{1}{2} B^2(x)\frac{\partial^2 g}{\partial x^2}\right] d\tau \\
    &+ B(x) \frac{\partial g}{\partial x} dw(\tau).
\end{align*}
where the first term results from the explicit time-dependence of $g(x,\tau)$.
Averaging over the stochastic variables $x(\tau)$ yields
\begin{align*}
    & \frac{d\overline{g(x,\tau)}}{d\tau}\\
    &= \int \!\! dx\, \left[-V(x) f + A(x) \frac{\partial f}{\partial x} + \frac{1}{2} B^2(x)\frac{\partial^2 f}{\partial x^2}\right] W(x,\tau)\\
    &= \int \!\! dx\, f(x)\left[-V(x) - \frac{\partial}{\partial x} A(x) + \frac{1}{2} \frac{\partial^2}{\partial x^2} B^2(x) \right] W(x,\tau).
\end{align*}
Using eq.~\eqref{eq:FPE-W} one finally arrives at
\begin{equation*}
    \frac{d\overline{g(x,\tau)}}{d\tau}= \int \!\! dx\, 
    f(x) \frac{\partial}{\partial \tau }W(x,\tau),
\end{equation*}
which proves eq.~\eqref{eq:average-generalized} from the main text.


\subsection{Sampling of the fully mixed state on AF Ising models}

\mfl{Due to the gauge freedom in the TWA} \tom{
a representation of the fully-mixed state can be achieved by several samplings in phase space. 
Here we show, how this non-uniqueness can be exploited to improve the approximation on AF Ising models also for short imaginary times.
}

\tom{
Starting with the marginal distribution $\tilde\chi_\theta$, the most common sampling of the fully mixed state is that one in which all \mfl{initial values $\theta_i$}  are uniformly distributed on the surface of the sphere, i.e. $\chi^{(i)}_{\theta}(0) \sim \sin\theta_i$. Note, since we are considering $\tilde\chi_\theta$, all spin states are represented by angles between $\arccos(1/\sqrt{3})$ and $\arccos(-1/\sqrt{3})$. \mfl{Equivalently,} it is possible to sample the fully mixed state by two equal \mfl{rings} on the upper and lower hemisphere, \mfl{e.g.} $\cos\theta_i = \pm\left(1 - \varepsilon \cdot u\right)/\sqrt{3}$, where $u$ is even\mfl{ly} distributed \mfl{in $[0,1)$, and} $\varepsilon\in[0,1]$ fixes the size of the two areas, and both cases $(\pm)$ appear with same probability.
}

\begin{figure}[h]
    \centering
        \includegraphics[width=0.48\textwidth]{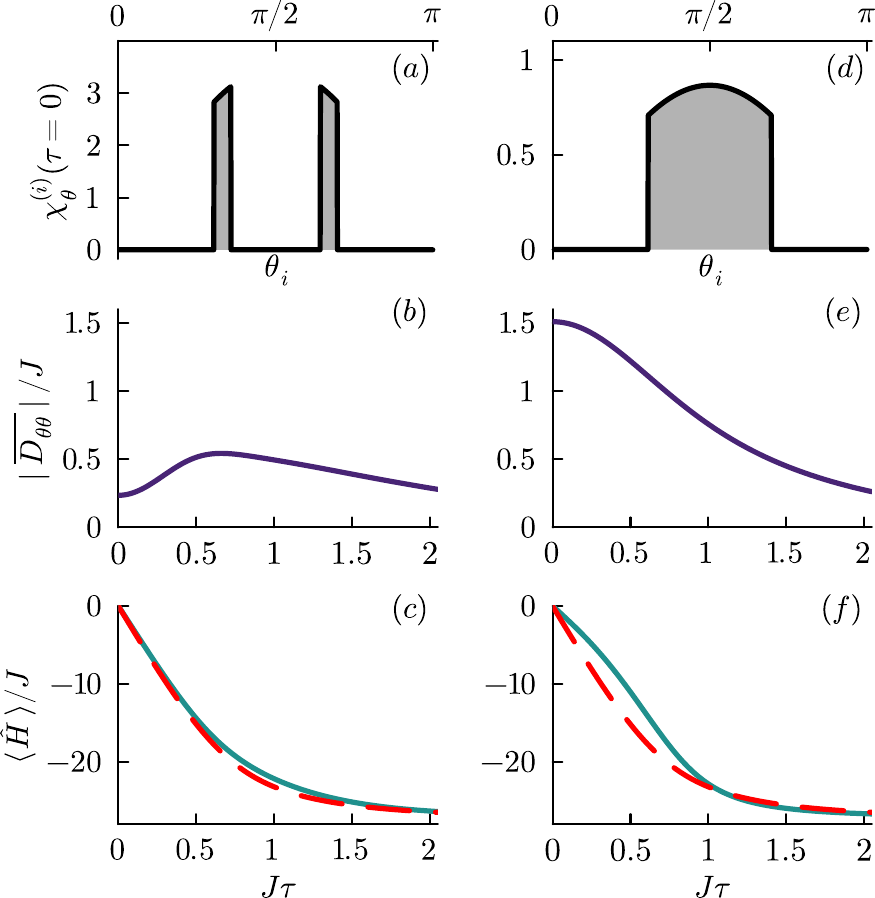}
        \caption{\tom{Effect of two different samplings for the fully-mixed state on the AF Ising model on the 3-regular graph in Fig.~\ref{fig:Ising-N22-3ru}(c). (a)+(d): Initial distributions $\chi_\theta^{(i)}(\tau = 0)$ for spin $(i)$ with $\varepsilon=0.25$ and $\varepsilon=1$, respectively. (b)+(e): Norm of the diffusion matrix $D_{\theta\theta}$, (c)+(f): Gibbs energy $\langle\hat H\rangle$ in $0\leq J\tau \leq 2$ for the initial samplings used in (a)+(d) and averaged over $N_\mathrm{traj}=10^5$ trajectories (iTWA: solid, green; ED: dashed, red).}}
        \label{fig:Different_Initial_Samplings}
\end{figure}

\tom{
In Fig.~\ref{fig:Different_Initial_Samplings} we compare \mfl{the} two initial samplings. (a)-(c) show the initial distribution for $\varepsilon=0.25$ as well as the norm of the diffusion matrix $\lvert \overline{D_{\theta\theta}}\rvert$ and the average energy $\langle \hat H \rangle$ for the graph simulated in Fig.~\ref{fig:Ising-N22-3ru}(c) in the main text. (d)-(f) show the same for the distribution over all spin states, i.e. $\varepsilon=1$. One \mfl{notices that} the sampling with $\varepsilon=0.25$ reduces the norm of the diffusion in $0 \leq J\tau\leq 2$ and the iTWA results \mfl{are} improved in this interval in comparison to the full sampling. The same sampling ($\varepsilon=0.25$) were chosen for the results in Fig.~\ref{fig:Ising-N22-3ru}.
}

\tom{
\mfl{Note that} the limiting case of $\varepsilon\to\ 0$, would \mfl{correspond to } two delta-distributions. \mfl{In this case} the entire dynamic disappears, \mfl{however}.} \mfl{Then one would have to sample exponentially many initial configurations to find the true ground state.}


\subsection{MaxCut on a random Erd\"os-Renyi graph}

\dennis{
In the main text we have \mfl{investiagted} the performance of iTWA solving Max-Cut on 3 regular graphs. We limited ourselves to this graph type because 3 regular graphs have classically known NP-/P-Hard regimes for approximation accuracy. To showcase the performance of iTWA on more general graphs, we \mfl{now discuss} Erd\"os-Renyi graphs with random edge weights and solve the Max-Cut problem on them using iTWA.
\\
We generate weighted Erd\"os-Renyi graphs ($G(N,p)$) by taking $N$ nodes, connecting each node to another with probability $p$, and \mfl{assign} a weight $rJ$ \mfl{to} each edge, where $r$ is a random number from a uniformly distributed interval $[0,1)$. Since keeping $p$ constant would lead to an all-to-all connected graph in the limit $N\rightarrow\infty$, we instead keep the mean number of edges $\langle k\rangle=Np$ constant.
\\
For the system size up to $N=100$ we can again utilize the GUROBI solver to find the Max-Cut of these Erd\"os-Renyi graphs and use it to benchmark the iTWA. Our metric here is \mfl{again} the relative energy deviation of the iTWA energy from the true ground energy}

\begin{align*}
    \Delta \epsilon=\frac{E_0 - E_\text{iTWA}(\tau,N_\text{traj})}{E_0}.
\end{align*}

\begin{figure}[h]
    \centering
        \includegraphics[width=0.48\textwidth]{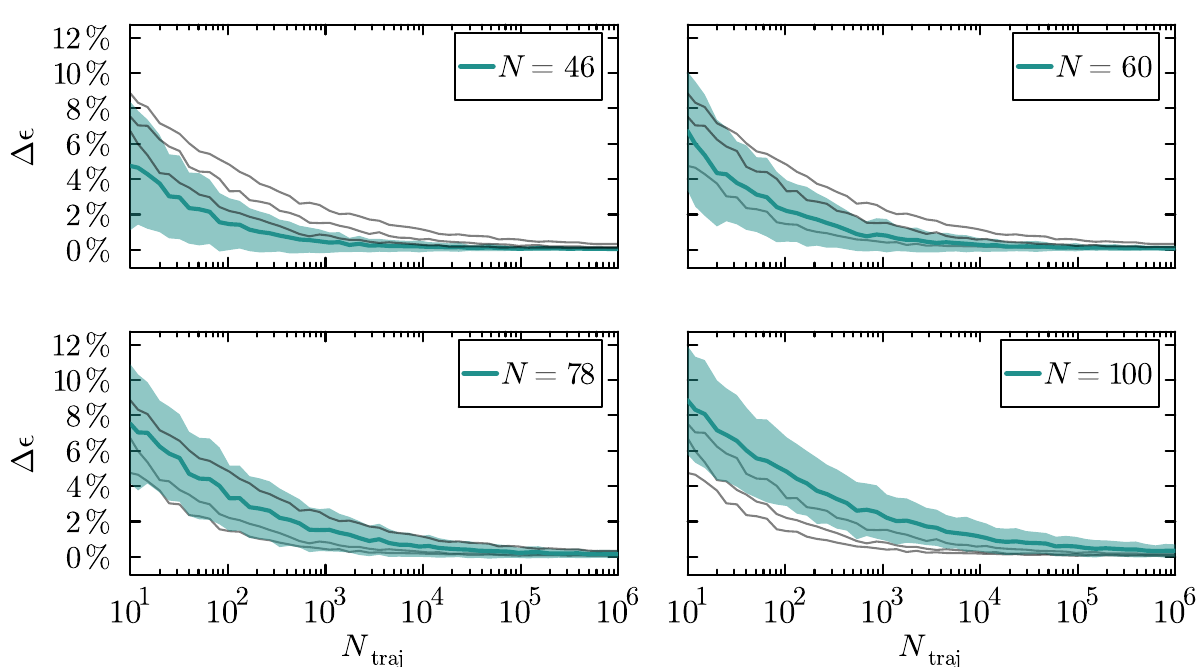}
        \caption{
        \tom{Mean relative error $\Delta\epsilon$ of the average energy  simulated via iTWA at an annealing time of $J\tau=10$ to the exact ground state energy estimated by Gurobi. The data are averaged over 200 randomly chosen Erd\"os-Renyi graphs with average degree $\langle k\rangle = Np=25$. Error bars describe the standard deviation. Again, the underlying distribution is lower bounded by $E_0$ for all trajectories.}}
        \label{fig:Ising-large-ErdosRenyi}
\end{figure}

\dennis{
In Fig.\ref{fig:Ising-large-ErdosRenyi} we can see \mfl{a} similar 
\mfl{behaviour as for } 3-regular graphs.  The \mfl{number of} trajectories $N_\text{traj}$ needed to converge increase with the system size $N$. This shows that \mfl{Ising models on} arbitrary graphs can be well approximated with the iTWA method. }


\subsection{Derivation of stochastic differential equations for the transverse-field Ising model}

\tom{
To derive useful SDEs for the transverse-field Ising model, we exploit the non-uniqueness of the correspondence rules. To this end, it turns out to be beneficial to shift the energy of the Hamiltonian, eq.~\eqref{eq:Transversse-field-Ising}, by a constant $-dJ$ with dimension $d$ of the lattice and to use a different correspondence rule for the unity matrix.
This has the advantage that the total $\phi$-dynamic is captured by up to 2nd order terms and can \mfl{be} fully account\mfl{ed} for in the simulations.}
Starting from
\begin{equation*}
    \label{eq:TFIM+Unitys}
    \hat H = - h\sum_j\hat \sigma_j^x - \frac{J}{2} \sum_{\langle ij\rangle } \hat\sigma_i^z\, \hat \sigma_j^z - dJ \sum_j \hat{\mathbbm{1}}_j
\end{equation*}
we can use the set $\left\{ \hat \Delta,\partial_\theta \hat \Delta, \partial_\phi \hat \Delta, \partial^2_\theta \hat \Delta \right\}$ to \tom{translate} the transverse-field and the Ising-coupling terms in correspondence to eq.~\eqref{eq:Density-EOM} as well as the set $\left\{\hat\Delta + \partial^2_\phi \hat \Delta,\partial_\theta \hat \Delta, \partial_\phi \hat \Delta, \partial^2_\theta \hat \Delta \right\}$ to translate the unitary matrices $\hat{\mathbbm{1}}_j$. \tom{The latter shifts the $D_{\phi\phi}$ matrix of the block diagonal diffusion matrix in the positive semi-definite region. The $D_{\theta\theta}$ block is not strictly positive, but the specific choice of correspondence rules for the terms in $\hat H$ makes it positive on average for certain initial states and sufficiently large values of $\tau$. After truncation of higher-order derivatives in $\theta$ we end up with the following drift vectors and diffusion matrices. The drift vectors read
\begin{eqnarray*}
    A_{\theta_i} &=&  \Big[ -J\sin\theta_i \sum_{\langle j\rangle} \cos\theta_j + \frac{h}{\sqrt{3}}\cos\theta_i\cos\phi_i
    \\
    &&+\ dJ\sin\theta_i\cos\theta_i \Big],
    \\
    A_{\phi_i} &=& -\frac{h}{\sqrt{3}}\csc\theta_i\sin\phi_i.
\end{eqnarray*}
}
Since the $D_{\theta\theta}$ block is \tom{in general} not positive, we additionally truncate the off-diagonal components $D_{\theta_i\theta_j}$ for~$\langle ij \rangle$. \tom{Finally,} the Wiener processes $B_{ij}$ are given by the square root of the diagonal elements of 
$D_{\theta\theta}$ for positive arguments (otherwise they are also truncated)
\begin{eqnarray*}
    B_{\theta_i\theta_i} &=& \Bigl[ 4J\cos\theta_i\sum_{j\in \langle ij\rangle}\cos\theta_j + \frac{4 h}{\sqrt{3}}\sin\theta_i\cos\phi_i 
    \\
    &&- 2dJ\sin^2\theta_i\Bigr]^{1/2}.
\end{eqnarray*}
and a decomposition of the matrix $\sf D_{\phi\phi}=\sf B_{\phi\phi}\sf B^\top_{\phi\phi}$, which exists because $D_{\phi\phi}$ is a positive \textit{Toeplitz} matrix
\begin{align}
    D_{\phi_i\phi_i} = 2dJ, && D_{\phi_i\phi_j} = -J \quad \textrm{for}\quad \langle ij \rangle.\nonumber
\end{align}
With this set of \tom{SDEs}, the results for the QPT in Fig.~\ref{fig:QPT} were achieved.

\end{document}